\documentclass[aps,prl,reprint,superscriptaddress,longbibliography]{revtex4-2}

\usepackage{graphicx}%
\usepackage{multirow}%
\usepackage{amsmath,amssymb,amsfonts}%
\usepackage{amsthm}%
\usepackage{mathrsfs}%

\usepackage{hyperref}
\usepackage[scaled=1.075,ncf,vvarbb]{newtxmath}
\usepackage{mathrsfs}
\usepackage{siunitx}
\usepackage{gensymb}
\usepackage{setspace}
\usepackage{tablefootnote}
\usepackage{todonotes}
\setuptodonotes{inline}  
\usepackage[normalem]{ulem}
\usepackage{comment}

\newcommand{\mrm}{\mathrm}

\newcommand{\ket}[1]{{\left| {#1} \right\rangle}}

\newcommand{\Esca}{\mathcal{E}}

\newcommand{\Bsca}{\mathcal{B}}

\begin{document}

\title{Sensing T-violating nuclear moments of paramagnetic ions in crystals}
\author{Aleksandar Radak}
\author{Mingyu Fan}
\author{Bassam Nima}
\affiliation{Department of Physics, University of Toronto, Toronto ON M5S 1A7, Canada}
\author{Yuiki Takahashi}
\affiliation{Division of Physics, Mathematics, and Astronomy, California Institute of Technology, Pasadena, California 91125 USA}
\author{Amar Vutha}
\affiliation{Department of Physics, University of Toronto, Toronto ON M5S 1A7, Canada}
\date{\today}

\begin{abstract}
Precision measurements of time-reversal (T) symmetry violating moments probe physics beyond the Standard Model. We show that precision spectroscopy of paramagnetic lanthanide and actinide ions doped into noncentrosymmetric crystals offers a promising platform for extending the sensitivity of searches for T-violation in nuclear physics. The unpaired valence electron in these ions allows the engineering of highly-coherent hyperfine transitions that are insensitive to magnetic fields, yet highly sensitive to new physics. These systems also provide other advantages for new physics searches, including large numbers of ions that can be measured in doped crystals, strong electric polarization of the ions by the crystal fields, enhancement of T-violating nuclear moments in nonspherical nuclei, and accurate comagnetometers generated by crystal symmetry. We estimate the new physics sensitivity of these proposed systems to be two orders of magnitude better than existing constraints.
\end{abstract}
\maketitle

The Standard Model of particle physics is known to be incomplete. In particular, new sources of time-reversal (T) symmetry violation are needed to explain the abundance of matter over antimatter in the observable universe. Precision measurements of parity (P)-odd T-odd moments in atomic and molecular systems have emerged as powerful probes for physics beyond the Standard Model, providing access to new physics at energy scales exceeding those in collider experiments \cite{demille_probing_2017}.

Atomic ions in crystals, whose internal quantum states can be optically controlled using lasers, are attracting interest as a promising new platform for T-violation measurements \cite{singh_new_2019, flambaum_electric_2020, ramachandran_nuclear_2023, morris_rare_2024}. 
These solid-state systems are particularly amenable for measuring symmetry-violating nuclear moments: the nuclei are significantly shielded against perturbations from the crystal lattice which enables measurements with long coherence times, whereas optical control and readout of nuclear spins allow high-precision symmetry-resolved measurements using large ensembles of nuclei.

In a crystal with noncentrosymmetric doping sites, such as yttrium orthosilicate (YSO), the dopant ions are in positions where parity is locally broken. The ions can therefore have large static electric polarization at equilibrium. The electrons in such a polarized ion create field gradients that interact with T-violating nuclear moments, leading to measurable P-odd T-odd perturbations in the hyperfine structure of the ion \cite{ramachandran_nuclear_2023}. The overall crystal symmetry, on the other hand, produces sub-ensembles of ions that are polarized in opposite directions. These ion sub-ensembles exhibit identical sensitivities to magnetic fields, but opposite sensitivities to T-violating effects. Thus, comparisons between these ``comagnetometer'' ensembles allow experiments to cleanly isolate new physics signals from spurious effects produced by magnetic fields, as demonstrated in Refs.\ \cite{nima_precision_2025,fan_wideband_2026}.

Lanthanide and actinide ions are ideal dopants in such crystals. They have valence electrons in $f$ orbitals that are shielded from the lattice by outer $s$ or $p$ electrons, so that the optical transitions of the valence electrons are only weakly perturbed by the crystal fields. As a result, narrow optical transitions with long coherence times are observed, which can be used to achieve efficient preparation and low-noise readout of hyperfine states \cite{macfarlane_coherent_1987}. Furthermore, some lanthanides and actinides have isotopes with significant nuclear nonsphericity, which enhances the sensitivity of these isotopes to hadronic T-violation \cite{flambaum_electric_2020,flambaum_enhanced_2022}.

Experiments using {diamagnetic} lanthanide ions in noncentrosymmetric crystals have previously been used to perform sensitive searches for new physics \cite{fan_wideband_2026}. Here, we show that hyperfine transitions  in \textit{paramagnetic} ions can be engineered to become insensitive to magnetic fields, yet retain high sensitivity to new physics.
We propose $^{167}$Er$^{3+}$, $^{229}$Th$^{3+}$, $^{233}$U$^{3+}$, and $^{235}$U$^{3+}$ ions doped in YSO as representative candidate systems for measuring P-odd T-odd nuclear moments, such as the nuclear Schiff moment (NSM) and the nuclear magnetic quadrupole moment (MQM). The above isotopes exhibit collective octupole or quadrupole modes that enhance their sensitivity to new physics \cite{dalton_enhanced_2023}. Optical transitions in these ions can be used for efficient state preparation and readout as in the diamagnetic case. Importantly, in addition, these systems can be engineered to have magnetic-field-insensitive hyperfine transitions. 

We project that a 10-day measurement with such systems can probe T-violating physics at a statistical sensitivity that is significantly better than current T-violating parameter limits, probing new particles at an energy scale $\sim100$ TeV. In addition, the same measurement can also set improved broadband bounds on axionlike particle (ALP) dark matter.
Outside of searches for nuclear T-violation, such transitions in, e.g., $^{229}$Th doped solids, may be useful for applications such as timekeeping and searches for the variability of fundamental constants.

\begin{figure}
    \centering
    \includegraphics[width=1\linewidth]{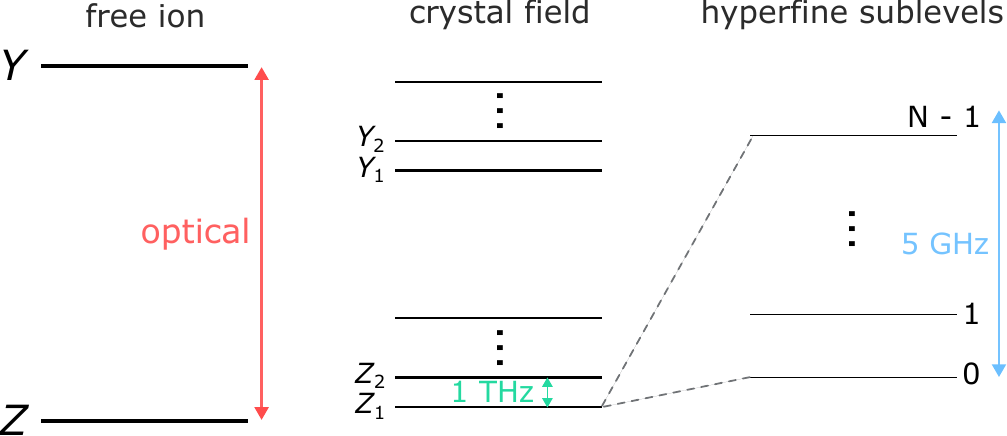}
    \caption{Schematic of the energy level structure of paramagnetic ions in YSO. The degeneracy of electron spin sublevels in the $Z$ and $Y$ electronic states is lifted by the interactions with the crystal. An optical transition between the two electronic states can be used for controlling and probing the ions. Each of the energy levels $Z_i, Y_j$ are doubly-degenerate Kramers pairs. Within each of these is a manifold of hyperfine states.}
    \label{fig:Energy levels}
\end{figure}

\textit{Hyperfine structure ---} When lanthanide or actinide atoms are doped into a YSO crystal, they substitute for yttrium ions and typically stabilize in a trivalent charge state. The effective Hamiltonian for the nuclear and electronic degrees of freedom in these ions can be written as \begin{equation}
    H = H_\mathrm{FI} + H_\mathrm{CF} + H_\mathrm{HF},
\end{equation} where $H_\mathrm{FI}$ is the free ion Hamiltonian, $H_\mathrm{CF}$ represents the interaction with the crystal field, and $H_\mathrm{HF}$ contains hyperfine and magnetic interactions  \cite{macfarlane_coherent_1987}. The energy scales involved in this Hamiltonian are schematically illustrated in Fig. \ref{fig:Energy levels}. Electronic transitions between eigenstates of the free ion typically occur at optical frequencies. Within each free ion electronic state, the interaction with the crystal breaks rotational symmetry and splits each electronic state by lifting the degeneracy of its magnetic levels. Each electronic state of paramagnetic systems, with a total electronic angular momentum of a half integer, forms fine-structure Kramers pairs under the influence of $H_\mrm{CF}$ \cite{klein_degeneracy_1952}. In the ground electronic state of the ion, denoted as $Z$, the fine-structure Kramers pairs are denoted $Z_1, Z_2, \cdots$ in increasing order of energy. For an excited state of the ion, denoted as $Y$, the fine-structure Kramers pairs are labelled $Y_1, Y_2, \cdots$ in increasing order of energy. Typical transitions between adjacent Kramers pairs occur in the THz frequency range.

At cryogenic temperatures (at $\lesssim 4$ K as typical in experiments), only the lowest electronic Kramers pair $Z_1$ is appreciably populated at thermal equilibrium. In this regime, the electronic degree of freedom can be described as an effective spin-1/2 system with the spin operator $\vec{S}$ \cite{afzelius_efficient_2010, tiranov_spectroscopic_2018}. The isotopes discussed here have half-integer nuclear spin ($I=5/2$ for $^{229}$Th and $^{233}$U, and $I=7/2$ for $^{167}$Er and $^{235}$U). Combining the electronic and nuclear degrees of freedom, there are $N = 2(2I+1)$ hyperfine states within the $Z_1$ manifold.

The hyperfine and Zeeman Hamiltonian for this manifold is \cite{wang_hyperfine_2023}
\begin{equation} \label{eq: hyperfine hamiltonian}
\begin{split}
    H_\mathrm{HF} & = A_{ij} S_i I_j + Q_{ij} I_i I_j  \\ & \ \ \ \ + \mu_B g_{ij} S_i \Bsca_j - \mu_N g_N \delta_{ij} I_i \Bsca_j.
\end{split}
\end{equation}
The indices $i,j \in \{x,y,z\}$ and repeated indices are summed over. The hyperfine coupling tensor is $A_{ij}$, the nuclear quadrupole interaction tensor is $Q_{ij}$, and $g_{ij}$ is the electronic g-tensor. Here $g_N$ is the nuclear g-factor, $\mu_B$ ($\mu_N$) is the Bohr (nuclear) magneton, and $\Bsca_j$ is the $j$-th component of the laboratory magnetic field. We denote eigenstates of $H_\mathrm{HF}$ as $\ket{0}, \cdots, \ket{N-1}$ in order of increasing energy. 

P-odd T-odd nuclear moments generated by physics beyond the Standard Model introduce further hyperfine interactions. The effective Hamiltonians describing the interaction of the nuclear Schiff moment (NSM) $\vec{\mathcal{S}}=\mathcal{S}\vec{I}/\lvert I\rvert$ and nuclear magnetic quadrupole moment (MQM) $\mathcal{M}$ \cite{flambaum_p-_1986, vutha_what_2026, flambaum_time-reversal_2014} are
\begin{equation}
    H_\mathrm{NSM} = \mathcal{W}_\mathcal{S}\mathcal{S} \frac{\vec{I}}{\left|\vec{I}\right|}\cdot\hat{n}
    \label{eq:schiff}
\end{equation}
and
\begin{equation}
    H_\mathrm{MQM} = - \mathcal{W}_\mathcal{M} \mathcal{M}\mathcal{I}_{ij} S_i n_j ,
    \label{eq:mqm}
\end{equation}
where $\mathcal{W}_\mathcal{S}$ ($\mathcal{W}_\mathcal{M}$) is a coefficient that characterizes the interaction of the NSM (MQM) with electrons in the ion. The unit vector $\hat{n}$ is the direction of electric polarization of the ion in the crystal. Here, $\mathcal{I}_{ij}$ is the rank 2 tensor operator $\mathcal{I}_{ij} =  \frac{1}{2I(2I-1)} \Big[ I_iI_j + I_jI_i  -  \frac{2}{3}\delta_{ij}I(I+1) \Big ]$.

In the YSO crystal ($C_2/c$ space group), there are two types of Y$^{3+}$ sites with different coordination numbers. We restrict our attention to the better-characterized ``site 1'' ions that have 7 nearest neighbor oxygen ions. Ions that substitute for Y$^{3+}$ in site 1 have four possible configurations that are related by two symmetries: a parity transformation (denoted as $\hat{\Pi}$), and a 180$^\circ$ rotation around the $b$ dielectric axis of the crystal (denoted as $\hat{\Sigma}$) \cite{nima_precision_2025}. The $\hat{\Pi}$ transformation relates two sub-ensembles of ions that have opposite electric polarization unit vector $\hat{n}$. We label these sub-ensembles of ions by the quantity $\pi \equiv \mathrm{sign}(\hat{n}\cdot \hat{x}) = \pm 1$, where we align the unit vectors in the lab-frame coordinate system ($\hat{x}$, $\hat{y}$, $\hat{z}$) with the dielectric axes of the crystal, $D_1$, $D_2$, and $b$, respectively.

As both $H_\mathrm{NSM}$ and $H_\mathrm{MQM}$ are dependent on the polarization direction $\hat{n}$, we can isolate P-odd T-odd energy shifts by measuring the \textit{difference} of hyperfine resonance frequencies between the $\pi = \pm 1$ sub-ensembles. In order to distinguish oppositely polarized ions, a lab electric field $\vec{\mathcal{E}}$ ($\sim$ 100 V/cm) can be applied to the crystal, which interacts with the atomic dipole moment difference between electronic states $\Delta D \hat{n}$ and generates an ion-polarization-dependent frequency shift $\Delta D \hat{n} \cdot \vec{\mathcal{E}}$. This optical frequency shift allows separately addressing the two sub-ensembles. Since the $\pi=\pm 1$ sub-ensembles have identical magnetic interactions but different new physics interactions, comparison between hyperfine resonances in these ions enables the separation of energy shifts due to new physics from those due to spurious magnetic fields, as demonstrated in Ref. \cite{nima_precision_2025}.

\textit{NTSC transitions ---} When lanthanide or actinide ions are doped in a crystal such as YSO, at special values of the lab magnetic field (called ``ZEFOZ points''), there exist pairs of hyperfine sublevels with equal magnetic moments along all three axes \cite{fraval_method_2004, rancic_coherence_2018, matsuura_exploration_2024}. A transition between such a pair of sublevels therefore becomes first-order insensitive to magnetic field fluctuations. In electron-spin-diamagnetic ions (such as Eu$^{3+}$ that was used in Refs.\ \cite{nima_precision_2025, fan_wideband_2026}), all ZEFOZ points coincide with zero sensitivity to new physics. Therefore there is no practical advantage to using such ions at ZEFOZ points for T-violation search experiments. Crucially, however, in \textit{electron-spin-paramagnetic} ions there exists a subset of ZEFOZ points where the system retains high sensitivity to P-odd T-odd new physics. We refer to these as nuclear T-violation-sensitive clock (NTSC) transitions.

The concept of engineering transitions that are insensitive to electromagnetic field fluctuations, but remain sensitive to T-violation, has been recently proposed \cite{takahashi_engineering_2023} and demonstrated in polar molecules \cite{takahashi_engineered_2025}. NTSC transitions for ions in crystals can be understood along similar lines. The essential idea is that the three terms in the hyperfine Hamiltonian -- the Zeeman, NSM, and MQM terms -- are each dependent on different combinations of $\vec{S}$ and $\vec{I}$. The Zeeman effect is dominated by $g_{ij} S_i \Bsca_j$, whereas the NSM depends on $\vec{I} \cdot \hat{n}$ and the MQM on $S_i n_j \mathcal{I}_{ij}$. Thus, two hyperfine states with identical magnetic moments must have nearly the same expectation values of electron spin projections along all axes; however, their nuclear spin vectors can have different orientations. Therefore, transitions between two such states can be insensitive to magnetic fields to first order, while remaining sensitive to the NSM and MQM. 

\begin{figure*}
    \centering
    \includegraphics[width=0.9\linewidth]{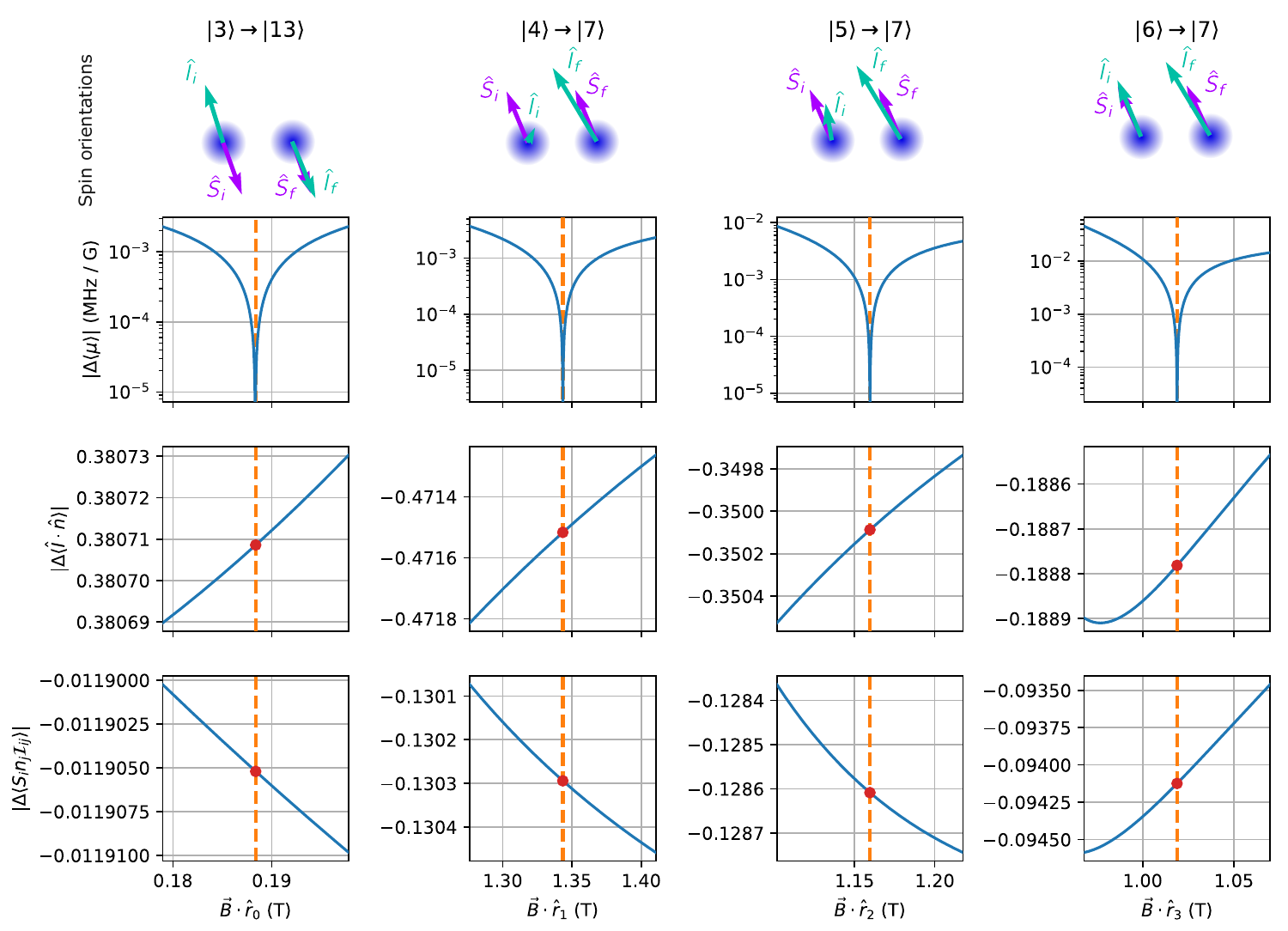}
    \caption{Example NTSC transitions for site 1 ions of $^{167}$Er:YSO. Each column represents a transition $\ket{i}\to\ket{f}$, with the hyperfine sublevels labeled in the order of low to high energy. The magnitude of the magnetic field is varied along the direction of the NTSC magnetic field for each transition ($\hat{r}_k$ for transitions labeled by $k\in 0, 1, 2, 3$). The top row shows the expectation values of the electron spin projection $\hat{S}=\vec{S} / |\vec{S}|$ and the nuclear spin projection $\hat{I}=\vec{I} / |\vec{I}|$ of initial and final states at the NTSC magnetic fields in the $x-y$ plane. The electron spins of the two states have nearly identical expectation values which ensures that both states have identical magnetic moments, while the nuclear spins in these states have different expectation values. The magnetic field sensitivity, the NSM sensitivity, and the MQM sensitivity are shown as a function of the magnetic field. The NTSC magnetic fields are labeled as orange dashed lines, and the NSM and MQM sensitivities at the NTSC magnetic field are denoted by red dots.}
    \label{fig:ntsc}
\end{figure*}

Using the hyperfine parameters of an electron-spin-paramagnetic ion, $^{167}$Er, located at site 1 in YSO \cite{wang_hyperfine_2023} as an example, we identify four NTSC transitions occurring at $|\Bsca| \lesssim$ 1 T that have high sensitivity to nuclear T-violation while also having appreciable transition dipole moments. These transitions are shown in Fig.\ \ref{fig:ntsc}. The coherence time of these transitions is estimated as approximately $\SI{0.1}{\second}$ (see the Appendix). Furthermore, different NTSC transitions have different sensitivities to the NSM and MQM, so that comparison of measured energy shifts across a set of NTSC transitions can measure the NSM and MQM individually. Additionally, at an NTSC point, all transitions other than the NTSC transition have large magnetic field sensitivities of $\sim10$ MHz / G. These transitions can therefore be used to detect and correct magnetic field imperfections that could generate systematic errors, as has been demonstrated in molecules \cite{takahashi_engineered_2025}. Combined with the comagnetometry method demonstrated in Ref.\ \cite{nima_precision_2025}, the overall experimental platform has extremely robust protection against systematic errors while maintaining high sensitivity to T-violation.

The existence of NTSC transitions is a generic and robust feature of electron-spin-paramagnetic ions in crystals such as YSO. To verify this fact, we numerically vary the magnitude of $Q_{ij}$ and $A_{ij}$ tensors of $^{167}$Er:YSO up and down by a factor of two. Over this entire parameter range, there exist NTSC transitions with comparable coherence time, as well as similar NSM and MQM sensitivities, as the transitions shown in Fig.\ \ref{fig:ntsc}. All the paramagnetic lanthanide and actinide ions considered here (Er$^{3+}$, Th$^{3+}$ and U$^{3+}$) have the same form of hyperfine Hamiltonian as Eq.~\ref{eq: hyperfine hamiltonian}, and thus they generically exhibit NTSC transitions.

\textit{Measurement protocol ---} To conduct precision searches of T-violating nuclear moments using these NTSC transitions, we propose a protocol with the essential idea similar to that for Eu:YSO \cite{nima_precision_2025}-- state preparation, radio-frequency (rf) spectroscopy, and comagnetometry readout. State preparation and detection rely on the optical transition between the ground $Z_1$ level and the excited $Y_1$ level. 

Similar to the measurement scheme used in Ref.\ \cite{fan_wideband_2026}, a chosen spectral class of ions is initially prepared in a hyperfine state $\ket{i}$ using spectral holeburning and optical pumping. Next, the hyperfine transition energy between $\ket{i}$ and a second sublevel $\ket{f}$ is probed using rf Ramsey spectroscopy. Finally, the population remaining in $\ket{i}$ is optically detected. 

During the detection, a laboratory electric field is applied to the crystal, which shifts the optical transition frequency in opposite directions for $\pi = \pm 1$ ions, as demonstrated in Ref.\ \cite{nima_precision_2025}. Thus the population remaining in $\ket{i}$ for each polarization of ion can be distinguished, allowing simultaneous measurements of the $\ket{i} \to \ket{f}$ transition frequency across the $\pi = \pm 1$ sub-ensembles. The sum of the two resonance frequencies measured in $\pi=\pm 1$ sub-ensembles only contains the Zeeman shift, whereas their difference isolates just the T-violating terms from the NSM and MQM. 

This measurement scheme can be applied to all the systems considered here (Er:YSO, Th:YSO and U:YSO). Er$^{3+}$ has a $^4I_{15/2}\to {}^4I_{13/2}$ transition at 1536 nm in YSO \cite{bottger_controlled_2008}. Th$^{3+}$ has a $^2F_{5/2}\to {}^2F_{7/2}$ optical transition at $\sim\SI{2}{\micro\meter}$ \cite{klinkenberg_spectrum_1949}. Optical spectroscopy of U$^{3+}$:LaCl$_3$ has identified the $^{4}I_{9/2}\to {}^4F_{3/2}$ transition at $\SI{1.4}{\micro\meter}$ \cite{karbowiak_spectral_2003}. All these intraconfigurational transitions are only weakly allowed via magnetic dipole ($M1$) transitions or through mixing effects of the crystal electric field, leading to narrow optical linewidths \cite{judd_optical_1962, ofelt_intensities_1962} that are convenient for hyperfine state-preparation and readout. 

The radioactive isotopes $^{229}$Th, $^{233}$U, and  $^{235}$U have half-lives of more than thousands of years, suitable to be incorporated in macroscopic crystals. Recently, optical excitation of $^{229}$Th ions in crystals has been achieved \cite{elwell_laser_2024, tiedau_laser_2024, zhang_frequency_2024}, and radioactive crystal production techniques have been demonstrated \cite{beeks_optical_2024}.  These developments confirm that radioactive dopants can be stably embedded in crystals for precision measurements, removing technical barriers to nuclear T-violation measurements with these nuclei. While neither Th or U have been doped into YSO, the isoelectronic lanthanides Ce and Nd have been doped into YSO and characterized \cite{alizadeh_intra-_2021, alizadeh_laser_2021}. Moreover, trivalent uranium-doped crystals have been produced for spectroscopy, including U:CaF$_2$ and U:LiYF$_4$ \cite{wittke_uranium-doped_1963, louis_reduction_1995}. Like YSO, these crystals may host other trivalent lanthanide ions, for instance, Pr, Eu, Ho, and Yb, indicating that U$^{3+}$ may be doped into YSO \cite{van_pieterson_4_2002, kumar_spectroscopic_2024, van_pieterson_4_2002, lyberis_effect_2012, sattler_electron-paramagnetic-resonance_1971, chiossi_optical_2024, singh_radiative_2023, lovric_spin_2012, lauritzen_spectroscopic_2012}. The two isotopes $^{233,235}$U have different intrinsic sensitivity to hadronic T-violation, and therefore comparing experimental measurements between these isotopes offers yet another means for stringent separation of systematic errors from genuine new physics effects.

\textit{Expected sensitivity ---} The attainable sensitivity of the proposed systems to T-violating physics can now be evaluated. The sensitivity of the NTSC transition frequency is 
\begin{equation}
    \Delta f = \frac{1}{2\pi T_\mathrm{coh} \mathrm{SNR}\sqrt{N}} = \SI{65}{\nano\hertz}.
    \label{eq:stat}
\end{equation}
In the above estimate we use a coherence time of $T_\mathrm{coh} = 0.14$ s, as calculated for the $\ket{4}\to\ket{7}$ NTSC transition at 1.343 T for $^{167}$Er:YSO (see the Appendix). We expect similar coherence times in crystals doped with Th and U isotopes. The signal-to-noise ratio $\mathrm{SNR} = 10^4$ is based on the estimated photon shot noise during spectroscopy (see the Appendix). We assume the sum of the state preparation and detection times to be equal to the coherence time, such that each measurement takes 0.28 s and $N=3\times10^6$ measurements can be made in ten days.

We use Eqs.~\ref{eq:schiff}~and~\ref{eq:mqm} to convert $\Delta f$ to the nuclear moment sensitivities. 
The electronic enhancement coefficients in YSO have been estimated to be an order of magnitude lower than those for polar molecules \cite{chen_relativistic_2024, flambaum_enhanced_2022, cheng_private_communication}, and therefore we assume reference values of $\mathcal{W}_\mathcal{S} \sim 100$ kHz/($e$ fm$^3$) and $\mathcal{W}_\mathcal{M} \sim 1$ MHz/($e$ fm$^2$). 

We use the NSM and MQM sensitivity factors, $\langle \hat{I}\cdot\hat{n}\rangle=-0.47$ and $\langle S_i n_j \mathcal{I}_{ij}\rangle=-0.13$ respectively, calculated for the $\ket{4} \to \ket{7}$ NTSC transition in $^{167}$Er:YSO. The statistical sensitivity of an experiment to hadronic T-violation effects can now be expressed in terms of the QCD $\theta$ parameter, which parametrizes the strength of a T-violating new physics term. These sensitivities are shown in Table~\ref{table:isotopes}. The calculated sensitivities of the proposed solid-state systems, $\delta\theta\sim10^{-12}$, exceed the current hadronic T-violation bound $\left|\theta\right|<10^{-10}$ set by the Hg and neutron electric dipole moment experiments \cite{graner_reduced_2016, abel_measurement_2020}. Furthermore, by analyzing temporal oscillations in $\theta$ derived from measurements of the NSM and MQM, we may search for ALP dark matter (e.g., \cite{fan_wideband_2026}). With the estimated experimental sensitivity to a possible NSM of $^{235}$U, we can constrain the ALP-gluon interaction strength between ALP masses $m_a$ of $5 \times 10^{-21}$ eV and $7 \times 10^{-15}$ eV at a sensitivity of 4 GeV$^{-1} \times m_a$ / (1 eV). This is an improvement of 2 to 4 orders of magnitudes over the previous limits set by T-violating moment experiments \cite{schulthess_new_2022, abel_search_2017, AxionLimits}.

\begin{table*}
    \centering
    \caption{Comparison of systems of different isotopes doped in YSO. Columns are species; Rows list nuclear spin $I$, the NSM $\mathcal{S}$ \cite{dalton_enhanced_2023}, the MQM $\mathcal{M}$ \cite{dalton_enhanced_2023}, and the estimated statistical sensitivities of the QCD $\theta$ parameter $\delta\theta_\mathrm{NSM}$ and  $\delta\theta_\mathrm{MQM}$, given the spectroscopy sensitivity estimated in Eq.~\ref{eq:stat}. 
    }
    \label{table:isotopes}
    \begin{ruledtabular}
        \begin{tabular}{cccccc}
            Isotope & $I$ & $\mathcal{S}$ ($\theta\times\mathrm{e}\cdot\mathrm{fm}^3$) & $\mathcal{M}$ ($\theta\times\mathrm{e}\cdot\mathrm{fm}^2$) & $\delta\theta_\mathrm{NSM}$ & $\delta\theta_\mathrm{MQM}$ \\
            \hline
            $^{167}$Er & 7/2 & -$^a$ & 0.13 & - & $\SI{4e-12}{}$\\
            $^{229}$Th & 5/2 & 1.2 & 0.09 & $\SI{1.2e-12}{}$ & $\SI{6e-12}{}$\\
            $^{233}$U & 5/2 & 0.44 & 0.09$^b$ & $\SI{3e-12}{}$ & $\SI{6e-12}{}$\\
            $^{235}$U & 7/2 & 1.9 & 0.09 & $\SI{7e-13}{}$ & $\SI{6e-12}{}$\\
            \hline
            \multicolumn{6}{l}{\footnotesize $^{a}$The NSM of $^{167}$Er is on the order of $\sim10^{-3}\theta\times\mathrm{e}\cdot\mathrm{fm}^3$ as it does not have octupolar enhancement, so it is not listed.}\\
            \multicolumn{6}{l}{\footnotesize $^{b}$For $^{233}$U we use the MQM of $^{235}$U since their masses are very similar.}

        \end{tabular}
    \end{ruledtabular}
\end{table*}

In conclusion, we summerize the advantages offered by paramagnetic lanthanide and actinide ions doped in noncentrosymmetric crystals for probing P-odd T-odd physics beyond the Standard Model. These systems combine nonspherical nuclei, with high intrinsic sensitivity to hadronic T-violating physics, together with noncentrosymmetric crystal sites that offer large enhancements of the nuclear moments generated by T-violation. The proposed systems possess robust means to reject systematic errors using (a) comagnetometers generated by crystal symmetries, and (b) magnetically-insensitive transitions that remain highly sensitive to T-violating new physics. We estimate that measurements of the nuclear Schiff moment and magnetic quadrupole moment using these solid-state systems can significantly improve the sensitivity of searches for nuclear T-violation, enabling tabletop-scale probes of new physics at the 100 TeV energy scale.

\begin{acknowledgments}
~\\
\textit{Acknowledgment.--} M.F. acknowledges funding from a CQIQC Postdoctoral Fellowship. Y.T. was supported by an IQIM Eddleman Fellowship and the Masason Foundation. This project was enabled by support from NSERC (grant no.\ SAPIN-2021-00025). 
\end{acknowledgments}

\appendix

\section{Appendix A: List of NTSC transitions in $^{167}$Er:YSO\label{appendix:ntsc}}

We discuss the new physics sensitivities of the calculated NTSC transitions in site 1 of $^{167}$Er:YSO as a prototypical example, since detailed spectroscopy data is available for this ion  \cite{wang_hyperfine_2023}. The time-varying magnetic field at the position of Er$^{3+}$ ions is dominated by fluctuations in the orientations of $^{89}$Y spins in the crystal, which reduce the coherence time of the transitions. As the linear Zeeman effect is absent, the coherence time of an NTSC transition at frequency $f_0$ is related to the quadratic Zeeman shift as \cite{zhong_optically_2015}
\begin{equation}
    \frac{1}{\pi T_2} = K_{ij} \, \Delta \Bsca_i \, \Delta \Bsca_j.
\end{equation}
Here $\Delta B \approx \SI{8}{\micro\tesla}$ is the estimated root-mean-squared magnetic field fluctuation due to the nearby Y nuclear spins \cite{zhong_optically_2015}, and $K_{ij} = \frac{\partial^2 f_0}{\partial \Bsca_i \partial \Bsca_j}$ is the curvature of the transition frequency with magnetic field. We assume that the magnetic field fluctuations are along the principal axis of $K_{ij}$ with the largest magnitude eigenvalue. We list the calculated coherence times of NTSC transitions in Table~\ref{table:zefoz} along with their NSM and MQM sensitivities.

\begin{table*}
    \centering
    \caption{Selected NTSC transition frequencies $f_0$, NTSC magnetic fields $B_0$, NSM and MQM sensitivities due to spin projection, transition magnetic dipole moment $\langle f| \mu_z| i \rangle$, largest eigenvalue magnitude of the quadratic Zeeman shift tensor $\lambda_\mathrm{max}(K)$, and calculated coherence time $T_2$ for the example of $^{167}$Er:YSO.}
    \label{table:zefoz}
    \begin{ruledtabular}
        \begin{tabular}{cccccccc}
            Transition & $f_0$ (MHz) & $B_0$ (T) & $\Delta \langle\hat{I}\cdot\hat{n}\rangle$ & $\Delta \langle S_i n_j \mathcal{I}_{ij}\rangle$ & $\langle f| \mu_z| i \rangle$ (MHz / T) & $\lambda_\mathrm{max}(K)$ (GHz / T$^2$) & $T_2$ (ms) \\
            \hline
            $\ket{2}\to\ket{12}$ & 3784 & 0.265 & -0.58 & -0.09 & 70 & 148 & 34 \\
            $\ket{3}\to\ket{13}$ & 3869 & 0.188 & 0.38 & -0.01 & 23 & 137 & 36 \\
            $\ket{4}\to\ket{7}$ & 2226 & 1.343 & -0.47 & -0.13 & 11 & 35 & 142 \\
            $\ket{4}\to\ket{12}$ & 3145 & 0.324 & 0.22 & -0.05 & 174 & 392 & 13 \\
            $\ket{5}\to\ket{7}$ & 1490 & 1.160 & -0.35 & -0.13 & 158 & 53 & 95 \\
            $\ket{6}\to\ket{7}$ & 754 & 1.019 & -0.19 & -0.09 & 361 & 139 & 36 \\
            
        \end{tabular}
    \end{ruledtabular}
\end{table*}

\section{Appendix B: Optical spectroscopy signal-to-noise ratio\label{appendix:snr}}

The $f-f$ transitions considered in this work are either $M1$ transitions or weakly allowed $E1$ transitions induced by the crystal field, with typical $E1$ oscillator strengths of $f_\mathrm{osc}\sim10^{-7}$ \cite{bottger_spectroscopy_2006}. With an incident laser of electric field amplitude $\Esca_0$, the Rabi frequency is $\Omega=d \Esca_0/\hbar$, where $d = \sqrt{3\hbar e^2 f_\mathrm{osc} / (2 m_e \omega_0)}$ is the transition electric dipole moment. Here $m_e$ is the electron mass and $\omega_0$ is the frequency of the optical transition.

Using realistic experimental parameters (transition wavelength $\lambda=c/(2\pi\omega_0) \approx \SI{1}{\micro\meter}$ and laser intensity $I=\epsilon_0 c\Esca_0^2/2=\SI{1}{\micro\watt\per\milli\meter\squared}$), we obtain
\begin{equation}
    \Omega = 2\pi\times640\, \mathrm{Hz}\sqrt{\frac{\lambda}{\SI{1}{\micro\meter}}\frac{I}{\SI{1}{\micro\watt\per\square\milli\meter}}\frac{10^{-7}}{f_\mathrm{osc}}}.
\end{equation}

To prevent excess optical pumping during spectroscopy, we excite less than $10\%$ of the population, resulting in a maximum optical detection time of $T_\mathrm{detect}\approx\SI{25}{\micro\second}$. With a crystal cross sectional area $A=\SI{1}{\milli\meter\squared}$, the number of photons detected is $N_\mathrm{photons}=I A \,T_\mathrm{detect} /(\hbar\omega_0)=1.3\times10^{8}$, giving a photon shot-noise-limited $\mathrm{SNR}=\sqrt{N_\mathrm{photons}}=1.1\times10^4$ for a spectral feature of optical depth $\sim1$.

\bibliographystyle{apsrev4-2}
\bibliography{references}

@unpublished{takahashi_engineered_2025,
      title={Engineered Molecular Clock Transitions for Symmetry Violation Searches}, 
      author={Yuiki Takahashi and Harish D. Ramachandran and Arian Jadbabaie and Yi Zeng and Chi Zhang and Nicholas R. Hutzler},
      year={2025},
      eprint={2508.06787},
      archivePrefix={arXiv},
      url={https://arxiv.org/abs/2508.06787}, 
}

@incollection{macfarlane_coherent_1987,
	title = {Coherent {Transient} and {Holeburning} {Spectroscopy} of {Rare} {Earth} {Ions} in {Solids}},
	volume = {21},
	isbn = {978-0-444-87051-3},
	url = {https://linkinghub.elsevier.com/retrieve/pii/B9780444870513500092},
	doi = {10.1016/B978-0-444-87051-3.50009-2},
	urldate = {2026-01-30},
	booktitle = {Modern {Problems} in {Condensed} {Matter} {Sciences}},
	publisher = {Elsevier},
	author = {Macfarlane, R.M. and Shelby, R.M.},
	year = {1987},
	pages = {51--184},
}

@article{wang_hyperfine_2023,
	title = {Hyperfine states of erbium doped yttrium orthosilicate for long-coherence-time quantum memories},
	volume = {262},
	issn = {00222313},
	url = {https://linkinghub.elsevier.com/retrieve/pii/S0022231323002685},
	doi = {10.1016/j.jlumin.2023.119935},
	urldate = {2026-01-30},
	journal = {J. Lumin.},
	author = {Wang, Shi-Jia and Chen, Yu-Hui and Longdell, Jevon J. and Zhang, Xiangdong},
	month = oct,
	year = {2023},
	pages = {119935},
}

@article{klein_degeneracy_1952,
	title = {On a {Degeneracy} {Theorem} of {Kramers}},
	volume = {20},
	issn = {0002-9505, 1943-2909},
	url = {https://pubs.aip.org/ajp/article/20/2/65/1034581/On-a-Degeneracy-Theorem-of-Kramers},
	doi = {10.1119/1.1933118},
	number = {2},
	urldate = {2026-01-30},
	journal = {Am. J. Phys.},
	author = {Klein, Martin J.},
	month = feb,
	year = {1952},
	pages = {65--71},
}

@article{zhang_frequency_2024,
	title = {Frequency ratio of the $^{229m}${Th} nuclear isomeric transition and the $^{87}${Sr} atomic clock},
	volume = {633},
	issn = {0028-0836, 1476-4687},
	url = {https://www.nature.com/articles/s41586-024-07839-6},
	doi = {10.1038/s41586-024-07839-6},
	number = {8028},
	urldate = {2026-01-30},
	journal = {Nature},
	author = {Zhang, Chuankun and Ooi, Tian and Higgins, Jacob S. and Doyle, Jack F. and Von Der Wense, Lars and Beeks, Kjeld and Leitner, Adrian and Kazakov, Georgy A. and Li, Peng and Thirolf, Peter G. and Schumm, Thorsten and Ye, Jun},
	month = sep,
	year = {2024},
	pages = {63--70},
}

@article{elwell_laser_2024,
	title = {Laser {Excitation} of the $^{229}${Th} {Nuclear} {Isomeric} {Transition} in a {Solid}-{State} {Host}},
	volume = {133},
	issn = {0031-9007, 1079-7114},
	url = {https://link.aps.org/doi/10.1103/PhysRevLett.133.013201},
	doi = {10.1103/PhysRevLett.133.013201},
	number = {1},
	urldate = {2026-01-30},
	journal = {Phys. Rev. Lett.},
	author = {Elwell, R. and Schneider, Christian and Jeet, Justin and Terhune, J. E. S. and Morgan, H. W. T. and Alexandrova, A. N. and Tran Tan, H. B. and Derevianko, Andrei and Hudson, Eric R.},
	month = jul,
	year = {2024},
	pages = {013201},
}

@article{tiedau_laser_2024,
	title = {Laser {Excitation} of the {Th}-229 {Nucleus}},
	volume = {132},
	issn = {0031-9007, 1079-7114},
	url = {https://link.aps.org/doi/10.1103/PhysRevLett.132.182501},
	doi = {10.1103/PhysRevLett.132.182501},
	number = {18},
	urldate = {2026-01-30},
	journal = {Phys. Rev. Lett.},
	author = {Tiedau, J. and Okhapkin, M. V. and Zhang, K. and Thielking, J. and Zitzer, G. and Peik, E. and Schaden, F. and Pronebner, T. and Morawetz, I. and De Col, L. Toscani and Schneider, F. and Leitner, A. and Pressler, M. and Kazakov, G. A. and Beeks, K. and Sikorsky, T. and Schumm, T.},
	month = apr,
	year = {2024},
	pages = {182501},
}

@article{nima_precision_2025,
	title = {Precision comagnetometry for {T}-violation searches in crystals},
	volume = {112},
	issn = {2469-9926, 2469-9934},
	url = {https://link.aps.org/doi/10.1103/jykv-rsd1},
	doi = {10.1103/jykv-rsd1},
	number = {3},
	urldate = {2026-01-29},
	journal = {Phys. Rev. A},
	author = {Nima, Bassam and Fan, Mingyu and Radak, Aleksandar and Jayich, Andrew M. and Vutha, Amar},
	month = sep,
	year = {2025},
	pages = {L030801},
}

@article{bottger_controlled_2008,
	title = {Controlled compositional disorder in {Er}$^{3+}$:{Y}$_2${Si}{O}$_5$ provides a wide-bandwidth spectral hole burning material at 1.5 $\upmu$m},
	volume = {77},
	copyright = {http://link.aps.org/licenses/aps-default-license},
	issn = {1098-0121, 1550-235X},
	shorttitle = {Controlled compositional disorder in {Er} 3 +},
	url = {https://link.aps.org/doi/10.1103/PhysRevB.77.155125},
	doi = {10.1103/PhysRevB.77.155125},
	number = {15},
	urldate = {2026-01-28},
	journal = {Phys. Rev. B},
	author = {Böttger, Thomas and Thiel, C. W. and Cone, R. L. and Sun, Y.},
	month = apr,
	year = {2008},
	pages = {155125},
}

@article{tiranov_spectroscopic_2018,
	title = {Spectroscopic study of hyperfine properties in $^{171}${Yb}$^{3+}$:{Y}$_2${SiO}$_5$},
	volume = {98},
	issn = {2469-9950, 2469-9969},
	shorttitle = {Spectroscopic study of hyperfine properties in {Yb} 3 + 171},
	url = {https://link.aps.org/doi/10.1103/PhysRevB.98.195110},
	doi = {10.1103/PhysRevB.98.195110},
	number = {19},
	urldate = {2026-01-28},
	journal = {Phys. Rev. B},
	author = {Tiranov, Alexey and Ortu, Antonio and Welinski, Sacha and Ferrier, Alban and Goldner, Philippe and Gisin, Nicolas and Afzelius, Mikael},
	month = nov,
	year = {2018},
	pages = {195110},
}

@article{afzelius_efficient_2010,
	title = {Efficient optical pumping of {Zeeman} spin levels in {Nd}$^{3+}$:{YVO}$_4$},
	volume = {130},
	copyright = {https://www.elsevier.com/tdm/userlicense/1.0/},
	issn = {00222313},
	url = {https://linkinghub.elsevier.com/retrieve/pii/S0022231309006280},
	doi = {10.1016/j.jlumin.2009.12.026},
	number = {9},
	urldate = {2026-01-28},
	journal = {J. Lumin.},
	author = {Afzelius, Mikael and Staudt, Matthias U. and De Riedmatten, Hugues and Gisin, Nicolas and Guillot-Noël, Olivier and Goldner, Philippe and Marino, Robert and Porcher, Pierre and Cavalli, Enrico and Bettinelli, Marco},
	month = sep,
	year = {2010},
	pages = {1566--1571},
}

@article{ramachandran_nuclear_2023,
	title = {Nuclear {T}-violation search using octopole-deformed nuclei in a crystal},
	volume = {108},
	url = {https://link.aps.org/doi/10.1103/PhysRevA.108.012819},
	doi = {10.1103/PhysRevA.108.012819},
	number = {1},
	journal = {Phys. Rev. A},
	author = {Ramachandran, H. D. and Vutha, A. C.},
	month = jul,
	year = {2023},
	pages = {012819},
}

@article{flambaum_enhanced_2022,
	title = {Enhanced magnetic quadrupole moments in nuclei with octupole deformation and their {C}{P}-violating effects in molecules},
	volume = {105},
	issn = {2469-9985, 2469-9993},
	url = {https://link.aps.org/doi/10.1103/PhysRevC.105.065503},
	doi = {10.1103/PhysRevC.105.065503},
	number = {6},
	urldate = {2026-01-27},
	journal = {Phys. Rev. C},
	author = {Flambaum, V. V. and Mansour, A. J.},
	month = jun,
	year = {2022},
	pages = {065503},
}

@article{flambaum_time-reversal_2014,
	title = {Time-{Reversal} {Symmetry} {Violation} in {Molecules} {Induced} by {Nuclear} {Magnetic} {Quadrupole} {Moments}},
	volume = {113},
	copyright = {http://link.aps.org/licenses/aps-default-license},
	issn = {0031-9007, 1079-7114},
	url = {https://link.aps.org/doi/10.1103/PhysRevLett.113.103003},
	doi = {10.1103/PhysRevLett.113.103003},
	number = {10},
	urldate = {2026-01-27},
	journal = {Phys. Rev. Lett.},
	author = {Flambaum, V. V. and DeMille, D. and Kozlov, M. G.},
	month = sep,
	year = {2014},
	pages = {103003},
}

@article{flambaum_electric_2020,
	title = {Electric dipole moments of atoms and molecules produced by enhanced nuclear {Schiff} moments},
	volume = {101},
	url = {https://link.aps.org/doi/10.1103/PhysRevA.101.042504},
	doi = {10.1103/PhysRevA.101.042504},
	number = {4},
	journal = {Phys. Rev. A},
	author = {Flambaum, V. V. and Dzuba, V. A.},
	month = apr,
	year = {2020},
	pages = {042504},
}

@article{demille_probing_2017,
	title = {Probing the frontiers of particle physics with tabletop-scale experiments},
	volume = {357},
	issn = {0036-8075, 1095-9203},
	url = {https://www.science.org/doi/10.1126/science.aal3003},
	doi = {10.1126/science.aal3003},
	pages = {990--994},
	number = {6355},
	journal = {Science},
	shortjournal = {Science},
	author = {{DeMille}, David and Doyle, John M. and Sushkov, Alexander O.},
	urldate = {2026-01-31},
	date = {2017-09-08},
	year = {2017},
}

@article{graner_reduced_2016,
	title = {Reduced {Limit} on the {Permanent} {Electric} {Dipole} {Moment} of ${}^{199}${Hg}},
	volume = {116},
	url = {https://link.aps.org/doi/10.1103/PhysRevLett.116.161601},
	doi = {10.1103/PhysRevLett.116.161601},
	number = {16},
	journal = {Phys. Rev. Lett.},
	author = {Graner, B. and Chen, Y. and Lindahl, E. G. and Heckel, B. R.},
	month = apr,
	year = {2016},
	pages = {161601},
}

@article{abel_measurement_2020,
	title = {Measurement of the {Permanent} {Electric} {Dipole} {Moment} of the {Neutron}},
	volume = {124},
	issn = {0031-9007, 1079-7114},
	url = {https://link.aps.org/doi/10.1103/PhysRevLett.124.081803},
	doi = {10.1103/PhysRevLett.124.081803},
	number = {8},
	urldate = {2024-08-12},
	journal = {Phys. Rev. Lett.},
	author = {Abel, C. and Afach, S. and Ayres, N. J. and Baker, C. A. and Ban, G. and Bison, G. and Bodek, K. and Bondar, V. and Burghoff, M. and Chanel, E. and Chowdhuri, Z. and Chiu, P.-J. and Clement, B. and Crawford, C. B. and Daum, M. and Emmenegger, S. and Ferraris-Bouchez, L. and Fertl, M. and Flaux, P. and Franke, B. and Fratangelo, A. and Geltenbort, P. and Green, K. and Griffith, W. C. and van der Grinten, M. and Grujić, Z. D. and Harris, P. G. and Hayen, L. and Heil, W. and Henneck, R. and Hélaine, V. and Hild, N. and Hodge, Z. and Horras, M. and Iaydjiev, P. and Ivanov, S. N. and Kasprzak, M. and Kermaidic, Y. and Kirch, K. and Knecht, A. and Knowles, P. and Koch, H.-C. and Koss, P. A. and Komposch, S. and Kozela, A. and Kraft, A. and Krempel, J. and Kuźniak, M. and Lauss, B. and Lefort, T. and Lemière, Y. and Leredde, A. and Mohanmurthy, P. and Mtchedlishvili, A. and Musgrave, M. and Naviliat-Cuncic, O. and Pais, D. and Piegsa, F. M. and Pierre, E. and Pignol, G. and Plonka-Spehr, C. and Prashanth, P. N. and Quéméner, G. and Rawlik, M. and Rebreyend, D. and Rienäcker, I. and Ries, D. and Roccia, S. and Rogel, G. and Rozpedzik, D. and Schnabel, A. and Schmidt-Wellenburg, P. and Severijns, N. and Shiers, D. and Tavakoli Dinani, R. and Thorne, J. A. and Virot, R. and Voigt, J. and Weis, A. and Wursten, E. and Wyszynski, G. and Zejma, J. and Zenner, J. and Zsigmond, G.},
	month = feb,
	year = {2020},
	pages = {081803},
}

@article{flambaum_p-_1986,
	title = {On the {P}- and {T}-nonconserving nuclear moments},
	volume = {449},
	copyright = {https://www.elsevier.com/tdm/userlicense/1.0/},
	issn = {03759474},
	url = {https://linkinghub.elsevier.com/retrieve/pii/0375947486903313},
	doi = {10.1016/0375-9474(86)90331-3},
	number = {4},
	urldate = {2026-02-02},
	journal = {Nucl. Phys. A},
	author = {Flambaum, V.V. and Khriplovich, I.B. and Sushkov, O.P.},
	month = mar,
	year = {1986},
	pages = {750--760},
}

@article{chen_relativistic_2024,
	title = {Relativistic {Exact} {Two}-{Component} {Coupled}-{Cluster} {Study} of {Molecular} {Sensitivity} {Factors} for {Nuclear} {Schiff} {Moments}},
	volume = {128},
	copyright = {https://doi.org/10.15223/policy-029},
	issn = {1089-5639, 1520-5215},
	url = {https://pubs.acs.org/doi/10.1021/acs.jpca.4c02640},
	doi = {10.1021/acs.jpca.4c02640},
	number = {31},
	urldate = {2026-02-02},
	journal = {J. Phys. Chem. A},
	author = {Chen, Tianxiang and Zhang, Chaoqun and Cheng, Lan and Ng, Kia Boon and Malbrunot-Ettenauer, Stephan and Flambaum, Victor V. and Lasner, Zack and Doyle, John M. and Yu, Phelan and Conn, Chandler J. and Zhang, Chi and Hutzler, Nicholas R. and Jayich, Andrew M. and Augenbraun, Benjamin and DeMille, David},
	month = aug,
	year = {2024},
	pages = {6540--6554},
}

@article{klinkenberg_spectrum_1949,
	title = {The spectrum of trebly ionized thorium, {Th} {IV}},
	volume = {15},
	copyright = {https://www.elsevier.com/tdm/userlicense/1.0/},
	issn = {00318914},
	url = {https://linkinghub.elsevier.com/retrieve/pii/0031891449900828},
	doi = {10.1016/0031-8914(49)90082-8},
	number = {8-9},
	urldate = {2026-02-02},
	journal = {Physica},
	author = {Klinkenberg, P.F.A. and Lang, R.J.},
	month = sep,
	year = {1949},
	pages = {774--788},
}

@article{alizadeh_intra-_2021,
	title = {Intra- and inter-configurational electronic transitions of {Ce}$^{3+}$-doped {Y}$_2${SiO}${_5}$ : {Spectroscopy} and crystal field analysis},
	volume = {117},
	issn = {09253467},
	shorttitle = {Intra- and inter-configurational electronic transitions of {Ce3}+-doped {Y2SiO5}},
	url = {https://linkinghub.elsevier.com/retrieve/pii/S0925346721003153},
	doi = {10.1016/j.optmat.2021.111114},
	urldate = {2026-02-02},
	journal = {Opt. Mater.},
	author = {Alizadeh, Y. and Martin, J.L.B. and Reid, M.F. and Wells, J.-P.R.},
	month = jul,
	year = {2021},
	pages = {111114},
}

@article{alizadeh_laser_2021,
	title = {Laser site-selective spectroscopy of {Nd}$^{3+}$-doped {Y}$_2${SiO}$_5$},
	volume = {234},
	issn = {00222313},
	url = {https://linkinghub.elsevier.com/retrieve/pii/S0022231321000752},
	doi = {10.1016/j.jlumin.2021.117959},
	urldate = {2026-03-15},
	journal = {J. Lumin.},
	author = {Alizadeh, Y. and Wells, J.-P.R. and Reid, M.F. and Ferrier, A. and Goldner, P.},
	month = jun,
	year = {2021},
	pages = {117959},
}

@article{lovric_spin_2012,
	title = {Spin {Hamiltonian} characterization and refinement for {Pr}$^{3+}$:{YAlO}$_3$ and {Pr}$^{3+}$:{Y}$_2${SiO}$_5$},
	volume = {85},
	copyright = {http://link.aps.org/licenses/aps-default-license},
	issn = {1098-0121, 1550-235X},
	shorttitle = {Spin {Hamiltonian} characterization and refinement for {Pr} 3 +},
	url = {https://link.aps.org/doi/10.1103/PhysRevB.85.014429},
	doi = {10.1103/PhysRevB.85.014429},
	number = {1},
	urldate = {2026-02-02},
	journal = {Phys. Rev. B},
	author = {Lovrić, Marko and Glasenapp, Philipp and Suter, Dieter},
	month = jan,
	year = {2012},
	pages = {014429},
}

@article{beeks_optical_2024,
	title = {Optical transmission enhancement of ionic crystals via superionic fluoride transfer: {Growing} {VUV}-transparent radioactive crystals},
	volume = {109},
	issn = {2469-9950, 2469-9969},
	shorttitle = {Optical transmission enhancement of ionic crystals via superionic fluoride transfer},
	url = {https://link.aps.org/doi/10.1103/PhysRevB.109.094111},
	doi = {10.1103/PhysRevB.109.094111},
	number = {9},
	urldate = {2026-02-02},
	journal = {Phys. Rev. B},
	author = {Beeks, Kjeld and Sikorsky, Tomas and Schaden, Fabian and Pressler, Martin and Schneider, Felix and Koch, Björn N. and Pronebner, Thomas and Werban, David and Hosseini, Niyusha and Kazakov, Georgy and Welch, Jan and Sterba, Johannes H. and Kraus, Florian and Schumm, Thorsten},
	month = mar,
	year = {2024},
	pages = {094111},
}

@article{rancic_coherence_2018,
	title = {Coherence time of over a second in a telecom-compatible quantum memory storage material},
	volume = {14},
	issn = {1745-2473, 1745-2481},
	url = {https://www.nature.com/articles/nphys4254},
	doi = {10.1038/nphys4254},
	number = {1},
	urldate = {2026-02-03},
	journal = {Nat. Phys.},
	author = {Rančić, Miloš and Hedges, Morgan P. and Ahlefeldt, Rose L. and Sellars, Matthew J.},
	month = jan,
	year = {2018},
	pages = {50--54},
}

@Unpublished{matsuura_exploration_2024,
	title = {Exploration of optimal hyperfine transitions for spin-wave storage in $^{167}${Er}$^{3+}$:{Y}$_2${SiO}$_5$},
	url = {http://arxiv.org/abs/2412.10126},
	doi = {10.48550/ARXIV.2412.10126},
    eprint = "2412.10126",
    archivePrefix = "arXiv",
	urldate = {2026-02-05},
	publisher = {arXiv},
	author = {Matsuura, K. and Yasui, S. and Kaji, R. and Sasakura, H. and Tawara, T. and Adachi, S.},
	year = {2024},
}

@article{zhong_optically_2015,
	title = {Optically addressable nuclear spins in a solid with a six-hour coherence time},
	volume = {517},
	issn = {0028-0836, 1476-4687},
	url = {https://www.nature.com/articles/nature14025},
	doi = {10.1038/nature14025},
	number = {7533},
	urldate = {2024-10-16},
	journal = {Nature},
	author = {Zhong, Manjin and Hedges, Morgan P. and Ahlefeldt, Rose L. and Bartholomew, John G. and Beavan, Sarah E. and Wittig, Sven M. and Longdell, Jevon J. and Sellars, Matthew J.},
	month = jan,
	year = {2015},
	pages = {177--180},
}

@article{fraval_method_2004,
  title = {Method of Extending Hyperfine Coherence Times in {Pr}$^{3+}$:{Y}$_{2}${SiO}$_{5}$},
  author = {Fraval, E. and Sellars, M. J. and Longdell, J. J.},
  journal = {Phys. Rev. Lett.},
  volume = {92},
  issue = {7},
  pages = {077601},
  numpages = {4},
  year = {2004},
  month = {Feb},
  publisher = {American Physical Society},
  doi = {10.1103/PhysRevLett.92.077601},
  url = {https://link.aps.org/doi/10.1103/PhysRevLett.92.077601}
}

@article{bottger_spectroscopy_2006,
	title = {Spectroscopy and dynamics of {Er}$^{3+}$:{Y}$_2${Si}{O}$_5$ at 1.5 $\upmu$m},
	volume = {74},
	copyright = {http://link.aps.org/licenses/aps-default-license},
	issn = {1098-0121, 1550-235X},
	shorttitle = {Spectroscopy and dynamics of {Er} 3 +},
	url = {https://link.aps.org/doi/10.1103/PhysRevB.74.075107},
	doi = {10.1103/PhysRevB.74.075107},
	number = {7},
	urldate = {2026-02-12},
	journal = {Phys. Rev. B},
	author = {Böttger, Thomas and Sun, Y. and Thiel, C. W. and Cone, R. L.},
	month = aug,
	year = {2006},
	pages = {075107},
}

@article{takahashi_engineering_2023,
	title = {Engineering {Field}-{Insensitive} {Molecular} {Clock} {Transitions} for {Symmetry} {Violation} {Searches}},
	volume = {131},
	issn = {0031-9007, 1079-7114},
	url = {https://link.aps.org/doi/10.1103/PhysRevLett.131.183003},
	doi = {10.1103/PhysRevLett.131.183003},
	number = {18},
	urldate = {2026-02-12},
	journal = {Phys. Rev. Lett.},
	author = {Takahashi, Yuiki and Zhang, Chi and Jadbabaie, Arian and Hutzler, Nicholas R.},
	month = oct,
	year = {2023},
	pages = {183003},
}

@Unpublished{vutha_what_2026,
	title = {What is a {Schiff} moment anyway?},
	url = {http://arxiv.org/abs/2601.07217},
	doi = {10.48550/arXiv.2601.07217},
    eprint = "2601.07217",
    archivePrefix = "arXiv",
	urldate = {2026-02-23},
	publisher = {arXiv},
	author = {Vutha, Amar},
	month = feb,
	year = {2026},
}

@article{ofelt_intensities_1962,
	title = {Intensities of {Crystal} {Spectra} of {Rare}-{Earth} {Ions}},
	volume = {37},
	issn = {0021-9606, 1089-7690},
	url = {https://pubs.aip.org/jcp/article/37/3/511/206596/Intensities-of-Crystal-Spectra-of-Rare-Earth-Ions},
	doi = {10.1063/1.1701366},
	number = {3},
	urldate = {2026-02-23},
	journal = {J. Chem. Phys.},
	author = {Ofelt, G. S.},
	month = aug,
	year = {1962},
	pages = {511--520},
}

@article{judd_optical_1962,
	title = {Optical {Absorption} {Intensities} of {Rare}-{Earth} {Ions}},
	volume = {127},
	copyright = {http://link.aps.org/licenses/aps-default-license},
	issn = {0031-899X},
	url = {https://link.aps.org/doi/10.1103/PhysRev.127.750},
	doi = {10.1103/PhysRev.127.750},
	number = {3},
	urldate = {2026-02-23},
	journal = {Phys. Rev.},
	author = {Judd, B. R.},
	month = aug,
	year = {1962},
	pages = {750--761},
}

@article{singh_new_2019,
	title = {A new concept for searching for time-reversal symmetry violation using {Pa}-229 ions trapped in optical crystals},
	volume = {240},
	issn = {0304-3843, 1572-9540},
	url = {http://link.springer.com/10.1007/s10751-019-1573-z},
	doi = {10.1007/s10751-019-1573-z},
	number = {1},
	urldate = {2024-09-20},
	journal = {Hyperfine Interactions},
	author = {Singh, Jaideep Taggart},
	month = dec,
	year = {2019},
	pages = {29},
}

@article{morris_rare_2024,
	title = {Rare isotope-containing diamond colour centres for fundamental symmetry tests},
	volume = {382},
	issn = {1364-503X, 1471-2962},
	url = {https://royalsocietypublishing.org/doi/10.1098/rsta.2023.0169},
	doi = {10.1098/rsta.2023.0169},
	number = {2265},
	urldate = {2026-02-23},
	journal = {Philos. Trans. R. Soc. A},
	author = {Morris, Ian M. and Klink, Kai and Singh, Jaideep T. and Mendoza-Cortes, Jose L. and Nicley, Shannon S. and Becker, Jonas N.},
	month = jan,
	year = {2024},
	pages = {20230169},
}

@article{karbowiak_spectral_2003,
	title = {Spectral intensities of {U}$^{\textrm{3+}}$ ions doped in {LaCl}$_{\textrm{3}}$ single crystals},
	volume = {101},
	issn = {0026-8976, 1362-3028},
	url = {http://www.tandfonline.com/doi/abs/10.1080/0026897021000046816},
	doi = {10.1080/0026897021000046816},
	number = {7},
	urldate = {2026-02-24},
	journal = {Mol. Phys.},
	author = {Karbowiak, Mirosław and Drożdżyński, Janusz},
	month = apr,
	year = {2003},
	pages = {971--975},
}

@misc{cheng_private_communication,
  author = "Cheng, Lan",
  howpublished = "private communication."
}

@article{dalton_enhanced_2023,
	title = {Enhanced {Schiff} and magnetic quadrupole moments in deformed nuclei and their connection to the search for axion dark matter},
	volume = {107},
	issn = {2469-9985, 2469-9993},
	url = {https://link.aps.org/doi/10.1103/PhysRevC.107.035502},
	doi = {10.1103/PhysRevC.107.035502},
	number = {3},
	urldate = {2026-03-03},
	journal = {Phys. Rev. C},
	author = {Dalton, F. and Flambaum, V. V. and Mansour, A. J.},
	month = mar,
	year = {2023},
	pages = {035502},
}

@article{wittke_uranium-doped_1963,
	title = {Uranium-doped calcium fluoride as a laser material},
	volume = {51},
	copyright = {https://ieeexplore.ieee.org/Xplorehelp/downloads/license-information/IEEE.html},
	issn = {0018-9219},
	url = {http://ieeexplore.ieee.org/document/1443589/},
	doi = {10.1109/PROC.1963.1659},
	number = {1},
	urldate = {2026-03-15},
	journal = {Proc. IEEE},
	author = {Wittke, J.P. and Kiss, Z.J. and Duncan, R.C. and McCormick, J.J.},
	year = {1963},
	pages = {56--62},
}

@article{louis_reduction_1995,
	title = {Reduction by $\gamma$-irradiation of tetravalent uranium in {LiYF$_4$}: {U} crystals for laser application},
	volume = {4},
	copyright = {https://www.elsevier.com/tdm/userlicense/1.0/},
	issn = {09253467},
	shorttitle = {Reduction by γ-irradiation of tetravalent uranium in {LiYF4}},
	url = {https://linkinghub.elsevier.com/retrieve/pii/092534679500002X},
	doi = {10.1016/0925-3467(95)00002-X},
	number = {5},
	urldate = {2026-03-15},
	journal = {Opt. Mater.},
	author = {Louis, M. and Simoni, E. and Hubert, S. and Gesland, J.Y.},
	month = jul,
	year = {1995},
	pages = {657--662},
}

@article{van_pieterson_4_2002,
	title = {$4f^n\rightarrow 4f^{n-1}5d$ transitions of the light lanthanides: {Experiment} and theory},
	volume = {65},
	copyright = {http://link.aps.org/licenses/aps-default-license},
	issn = {0163-1829, 1095-3795},
	shorttitle = {4 f n → 4 f n − 1 5 d transitions of the light lanthanides},
	url = {https://link.aps.org/doi/10.1103/PhysRevB.65.045113},
	doi = {10.1103/PhysRevB.65.045113},
	number = {4},
	urldate = {2026-03-15},
	journal = {Phys. Rev. B},
	author = {Van Pieterson, L. and Reid, M. F. and Wegh, R. T. and Soverna, S. and Meijerink, A.},
	month = jan,
	year = {2002},
	pages = {045113},
}

@article{kumar_spectroscopic_2024,
	title = {Spectroscopic studies of {Eu} doped {CaF$_2$} single crystal},
	issn = {22147853},
	url = {https://linkinghub.elsevier.com/retrieve/pii/S2214785324003183},
	doi = {10.1016/j.matpr.2024.05.020},
	urldate = {2026-03-15},
	journal = {Mater. Today: Proc.},
	author = {Kumar, Ravinder and Joseph, David},
	month = may,
	year = {2024},
	pages = {S2214785324003183},
}

@article{lyberis_effect_2012,
	title = {Effect of {Yb}$^{3+}$ concentration on optical properties of {Yb}:{CaF$_2$} transparent ceramics},
	volume = {34},
	copyright = {https://www.elsevier.com/tdm/userlicense/1.0/},
	issn = {09253467},
	shorttitle = {Effect of {Yb3}+ concentration on optical properties of {Yb}},
	url = {https://linkinghub.elsevier.com/retrieve/pii/S0925346711003107},
	doi = {10.1016/j.optmat.2011.05.036},
	number = {6},
	urldate = {2026-03-15},
	journal = {Opt. Mater.},
	author = {Lyberis, Andréas and Stevenson, Adam J. and Suganuma, Akiko and Ricaud, Sandrine and Druon, Frédéric and Herbst, Frédéric and Vivien, Daniel and Gredin, Patrick and Mortier, Michel},
	month = apr,
	year = {2012},
	pages = {965--968},
}

@article{sattler_electron-paramagnetic-resonance_1971,
	title = {Electron-{Paramagnetic}-{Resonance} {Spectra} of {Nd}$^{3+}$, {Dy}$^{3+}$, {Er}$^{3+}$, and {Yb}$^{3+}$ in {Lithium} {Yttrium} {Fluoride}},
	volume = {4},
	copyright = {http://link.aps.org/licenses/aps-default-license},
	issn = {0556-2805},
	url = {https://link.aps.org/doi/10.1103/PhysRevB.4.1},
	doi = {10.1103/PhysRevB.4.1},
	number = {1},
	urldate = {2026-03-15},
	journal = {Phys. Rev. B},
	author = {Sattler, J. P. and Nemarich, J.},
	month = jul,
	year = {1971},
	pages = {1--5},
}

@article{chiossi_optical_2024,
	title = {Optical coherence and spin population dynamics in $^{171}${Yb}$^{3+}$:{Y}$_2${SiO}$_5$ single crystals},
	volume = {109},
	issn = {2469-9950, 2469-9969},
	shorttitle = {Optical coherence and spin population dynamics in {Yb} 3 + 171},
	url = {https://link.aps.org/doi/10.1103/PhysRevB.109.094114},
	doi = {10.1103/PhysRevB.109.094114},
	number = {9},
	urldate = {2026-03-15},
	journal = {Phys. Rev. B},
	author = {Chiossi, Federico and Lafitte-Houssat, Eloïse and Ferrier, Alban and Welinski, Sacha and Morvan, Loïc and Berger, Perrine and Serrano, Diana and Afzelius, Mikael and Goldner, Philippe},
	month = mar,
	year = {2024},
	pages = {094114},
}

@article{singh_radiative_2023,
	title = {Radiative properties of green-emitting {Ho$^{3+}$} doped {Y$_2$SiO$_5$} system: exploring the potential use as a phosphor},
	volume = {34},
	issn = {0957-4522, 1573-482X},
	shorttitle = {Radiative properties of green-emitting {Ho3}+ doped {Y2SiO5} system},
	url = {https://link.springer.com/10.1007/s10854-023-11710-1},
	doi = {10.1007/s10854-023-11710-1},
	number = {36},
	urldate = {2026-03-15},
	journal = {J. Mater. Sci.: Mater. Electron.},
	author = {Singh, Vijay and Prasad, Aman and Seshadri, M. and Kaur, Sumandeep and Rao, A. S.},
	month = dec,
	year = {2023},
	pages = {2304},
}

@article{lauritzen_spectroscopic_2012,
	title = {Spectroscopic investigations of {Eu}$^{3+}$:{Y}$_2${SiO}$_5$ for quantum memory applications},
	volume = {85},
	copyright = {http://link.aps.org/licenses/aps-default-license},
	issn = {1098-0121, 1550-235X},
	shorttitle = {Spectroscopic investigations of {Eu} 3 +},
	url = {https://link.aps.org/doi/10.1103/PhysRevB.85.115111},
	doi = {10.1103/PhysRevB.85.115111},
	number = {11},
	urldate = {2024-07-17},
	journal = {Phys. Rev. B},
	author = {Lauritzen, B. and Timoney, N. and Gisin, N. and Afzelius, M. and De Riedmatten, H. and Sun, Y. and Macfarlane, R. M. and Cone, R. L.},
	month = mar,
	year = {2012},
	pages = {115111},
}

@article{fan_wideband_2026,
	title = {Wideband search for axionlike dark matter using octupolar nuclei in a crystal},
	issn = {0031-9007, 1079-7114},
	url = {https://link.aps.org/doi/10.1103/dm9j-9pry},
	doi = {10.1103/dm9j-9pry},
	urldate = {2026-03-15},
	journal = {Phys. Rev. Lett.},
	author = {Fan, Mingyu and Nima, Bassam and Radak, Aleksandar and Alonzo-Álvarez, Gonzalo and Vutha, Amar},
	month = mar,
	year = {2026},
}

@article{schulthess_new_2022,
	title = {New {Limit} on {Axionlike} {Dark} {Matter} {Using} {Cold} {Neutrons}},
	volume = {129},
	issn = {0031-9007, 1079-7114},
	url = {https://link.aps.org/doi/10.1103/PhysRevLett.129.191801},
	doi = {10.1103/PhysRevLett.129.191801},
	number = {19},
	urldate = {2026-03-17},
	journal = {Phys. Rev. Lett.},
	author = {Schulthess, Ivo and Chanel, Estelle and Fratangelo, Anastasio and Gottstein, Alexander and Gsponer, Andreas and Hodge, Zachary and Pistillo, Ciro and Ries, Dieter and Soldner, Torsten and Thorne, Jacob and Piegsa, Florian M.},
	month = nov,
	year = {2022},
	pages = {191801},
}

@article{abel_search_2017,
	title = {Search for {Axionlike} {Dark} {Matter} through {Nuclear} {Spin} {Precession} in {Electric} and {Magnetic} {Fields}},
	volume = {7},
	copyright = {https://creativecommons.org/licenses/by/4.0/},
	issn = {2160-3308},
	url = {https://link.aps.org/doi/10.1103/PhysRevX.7.041034},
	doi = {10.1103/PhysRevX.7.041034},
	number = {4},
	urldate = {2026-03-17},
	journal = {Phys. Rev. X},
	author = {Abel, C. and Ayres, N. J. and Ban, G. and Bison, G. and Bodek, K. and Bondar, V. and Daum, M. and Fairbairn, M. and Flambaum, V. V. and Geltenbort, P. and Green, K. and Griffith, W. C. and Van Der Grinten, M. and Grujić, Z. D. and Harris, P. G. and Hild, N. and Iaydjiev, P. and Ivanov, S. N. and Kasprzak, M. and Kermaidic, Y. and Kirch, K. and Koch, H.-C. and Komposch, S. and Koss, P. A. and Kozela, A. and Krempel, J. and Lauss, B. and Lefort, T. and Lemière, Y. and Marsh, D. J. E. and Mohanmurthy, P. and Mtchedlishvili, A. and Musgrave, M. and Piegsa, F. M. and Pignol, G. and Rawlik, M. and Rebreyend, D. and Ries, D. and Roccia, S. and Rozpędzik, D. and Schmidt-Wellenburg, P. and Severijns, N. and Shiers, D. and Stadnik, Y. V. and Weis, A. and Wursten, E. and Zejma, J. and Zsigmond, G.},
	month = nov,
	year = {2017},
	pages = {041034},}

@misc{AxionLimits,
  author       = {Ciaran O'Hare},
  title        = {cajohare/AxionLimits: AxionLimits},
  month        = jul,
  year         = 2020,
  publisher    = {Zenodo},
  version      = {v1.0},
  doi          = {10.5281/zenodo.3932430},
}

\end{document}